\documentclass[aps,prl,onecolumn,groupedaddress,showpacs,showkeys]{revtex4}
\usepackage{graphicx}
\usepackage{amsmath}

\newcommand{\ef}{E_{\scriptscriptstyle F}}
\newcommand{\kf}{k_{\scriptscriptstyle F}}

\newcommand{\leqa}{\stackrel{<}{\scriptstyle \sim}}

\newcommand{\rb}{\bar{\rho}}
\newcommand{\rt}{\widetilde{\rho}}
\newcommand{\Db}{\bar{\Delta}}
\newcommand{\Dt}{\widetilde{\Delta}}

\newcommand{\Ecb}{\bar{E}_{\rm C}}
\newcommand{\Ect}{\widetilde{E}_{\rm C}}
\newcommand{\rms}[1]{\sqrt{\left< #1^2 \right>}}
\newcommand{\secmom}[1]{\left< #1^2 \right>}
\newcommand{\tm}{\tau_{\rm min}}
\newcommand{\tD}{\tau_{\scriptscriptstyle \Delta}}
\newcommand{\tH}{\tau_{\scriptscriptstyle H}}
\newcommand{\eps}{\varepsilon}

\begin{document}

%\title{Mesoscopic Fluctuations of the Pairing Gap}{Mesoscopic Fluctuations of the Pairing Gap}

%\author{S. \AA berg}{address={Mathematical Physics, LTH, Lund University, P.O. Box 118, S-221 00 Lund, Sweden}}
%\author{H. Olofsson}{address={Mathematical Physics, LTH, Lund University, P.O. Box 118, S-221 00 Lund, Sweden}}
%\author{P. Leboeuf}{address={Laboratoire de Physique Th{\'e}orique et
%    Mod{\`e}les Statistiques, CNRS, B{\^a}t. 100, Universit{\'e} de Paris-Sud,
%    91405 Orsay Cedex, France}}

\title{Mesoscopic Fluctuations of the Pairing Gap}%

\author{S. \AA berg,$^1$, H. Olofsson,$^1$ and P. Leboeuf$~^2$}

\affiliation{$^1$Mathematical Physics, LTH, Lund
  University, P.O. Box 118, S-221 00 Lund, Sweden \\
$^2$Laboratoire de Physique Th{\'e}orique et Mod{\`e}les
Statistiques, CNRS, B{\^a}t. 100, Universit{\'e} de Paris-Sud, 91405
Orsay Cedex, France}

\begin{abstract}
A description of mesoscopic fluctuations of the pairing gap in
finite-sized quantum systems based on periodic orbit theory is
presented. The size of the fluctuations are found to depend on
quite general properties. We distinguish between systems
where corresponding classical motion is regular or chaotic, and
describe in detail fluctuations of the BCS gap as a function of
the size of the system. The theory is applied to different
mesoscopic systems: atomic nuclei, metallic grains, and ultracold
fermionic gases. We also present a detailed description of pairing
gap variation with particle number for nuclei based on a deformed
cavity potential.
\end{abstract}

\maketitle

\section{Introduction}
Bohr, Mottelson and Pines were first to apply the Bardeen, Cooper,
Schrieffer (BCS) theory of superconductivity to finite size
systems, namely to describe pairing in atomic nuclei \cite{Pines}.
A consequence of the finite size of the system is the appearance
of shell structure. This implies fluctuations of the pairing gap
as a parameter, like the particle number, is varied. In this
contribution we shall discuss how these fluctuations can be
described in a semiclassical theory.

We first discuss pairing in nuclei obtained from the odd-even mass
difference. Pairing gaps calculated from different mass models are
compared, both with respect to average gaps and to fluctuations.
In the next section a semiclassical theory of fluctuations of the
BCS pairing gap \cite{Olofsson} is presented. This provides
analytic expressions of fluctuations of the pairing gap, where the
dynamics of the underlying classical system (chaos/order) is an
important parameter. The theory is first applied to pairing
fluctuations in nuclei, where a detailed comparison with data is
performed. We also utilize explicit periodic orbits, taken from a
deformed cavity model, to describe the detailed variation of the
pairing gap with particle number. Nano-sized metallic grains are
also studied, where, due to a lack of experimental data, we
compare our theoretical results to other existing numerical
results. Interaction strength, external potential, as well as the
number of particles can be experimentally tuned in ultracold
fermionic gases, and we discuss the size of pairing fluctuations
for such systems, as obtained from our theory.

\section{Odd-even mass difference in nuclei}
\begin{figure}
\includegraphics[width=10cm,clip=true]{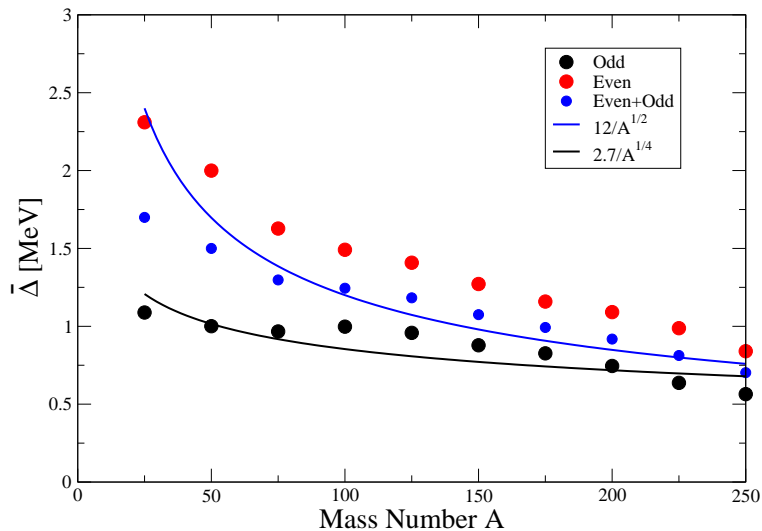}
\vspace{-0.3cm}
 \caption{Average pairing gap, $\Delta$, versus particle number $A$, as
 obtained from Eq.~(\ref{D3}) using measured masses
 \cite{NuclMass}. The averaging is performed in a region of fixed mass number, $A$,
 over all available isotopes.
 The particle number $M$ (neutron or proton number)
 is either odd (lower black dots), even (upper red dots), or both even and odd
 (middle blue dots). The curves, $2.7/A^{1/4}$~MeV (lower
 line) and $12/A^{1/2}$~MeV (upper line), are obtained from
 fits to cases with odd $M$, and both odd and even $M$-values,
 respectively.}
 \label{Fig1}
\end{figure}

The systematic difference between the ground-state mass of odd and
even nuclei constitutes an important indicator of pairing in
nuclei. The pairing gap can be calculated from binding energies,
$B$, utilizing the three--point measure
\begin{equation}
\Delta_3 (M) = B(M) - \frac{1}{2}[B(M+1) + B(M-1)],
 \label{D3}
\end{equation}
where $M$ is the neutron $N$ or proton $Z$ number. In the presence
of other possible interactions, this quantity has been shown to be
a good measure of pairing correlations \cite{doba} provided $M$ is
taken as an odd number. In that case, it is easy to see that there
is no contribution from the mean field in $\Delta_3$, while if $M$
is taken as an even number, an extra contribution to $\Delta_3$
from the single-particle levels appears in the extreme
single-particle model as $\frac{1}{2}(e_{i+1}-e_i)$, where $e_i$
is the last occupied single-particle level. In Fig.~\ref{Fig1} we
see the systematic difference in the average of $\Delta_3$ when
$M$ is an odd and even number. Restricting to $\Delta_3$-values
obtained from odd numbers of $M$, a fit of the pairing gap gives
\begin{equation}
  \Db = \frac{2.7}{A^{1/4}}~{\rm MeV} \ ,
  \label{Dbeq}
\end{equation}
where $A=N+Z$ is the total number of nucleons. If, however, also
cases with $M$ is an even number are included, the usually
employed pairing gap value,
\begin{equation}
 \Db=\frac{12}{A^{1/2}}~{\rm MeV} \ ,
 \label{Dbeqold}
\end{equation}
gives a better fit. The fitted difference $\Delta_3$(even
M)-$\Delta_3$(odd M) = $\frac{1}{2}(e_{i+1}-e_i)$ is found to be
inversely proportional to the mass number, $A$, and vary as $50/A$
MeV.
\begin{figure}
\includegraphics[width=10cm,clip=true]{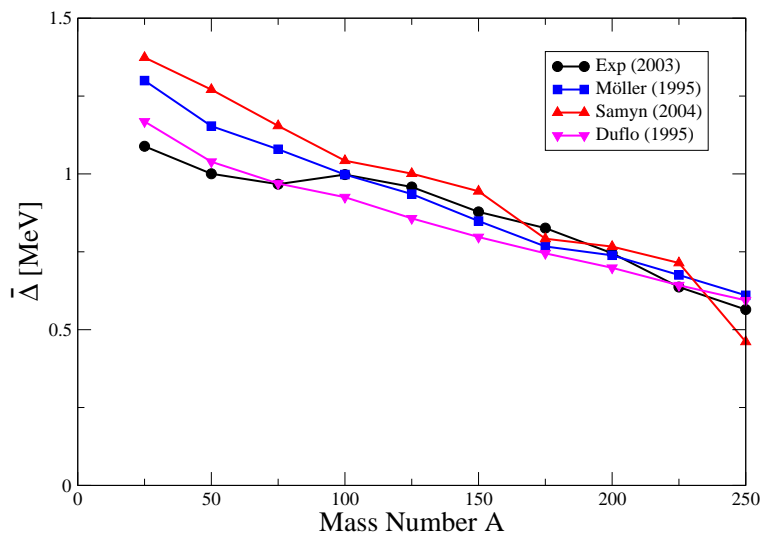}
\vspace{-0.3cm}
 \caption{Average pairing gap versus mass number $A$ using the
 three-point measure, Eq.~(\ref{D3}), with odd $M$. Binding
 energies are obtained from
 different theoretical mass models, M\"{o}ller et al \cite{Moller} (black
 dots), Samyn et al (red triangles pointing up) \cite{Samyn} and Duflo et al
 (magenta triangles pointing down) \cite{Duflo}. The theoretical curves are
 compared to experimental results (black dots), cf Fig.~~\ref{Fig1}.}
 \label{Fig2}
\end{figure}
\begin{figure}
\includegraphics[width=14cm,clip=true]{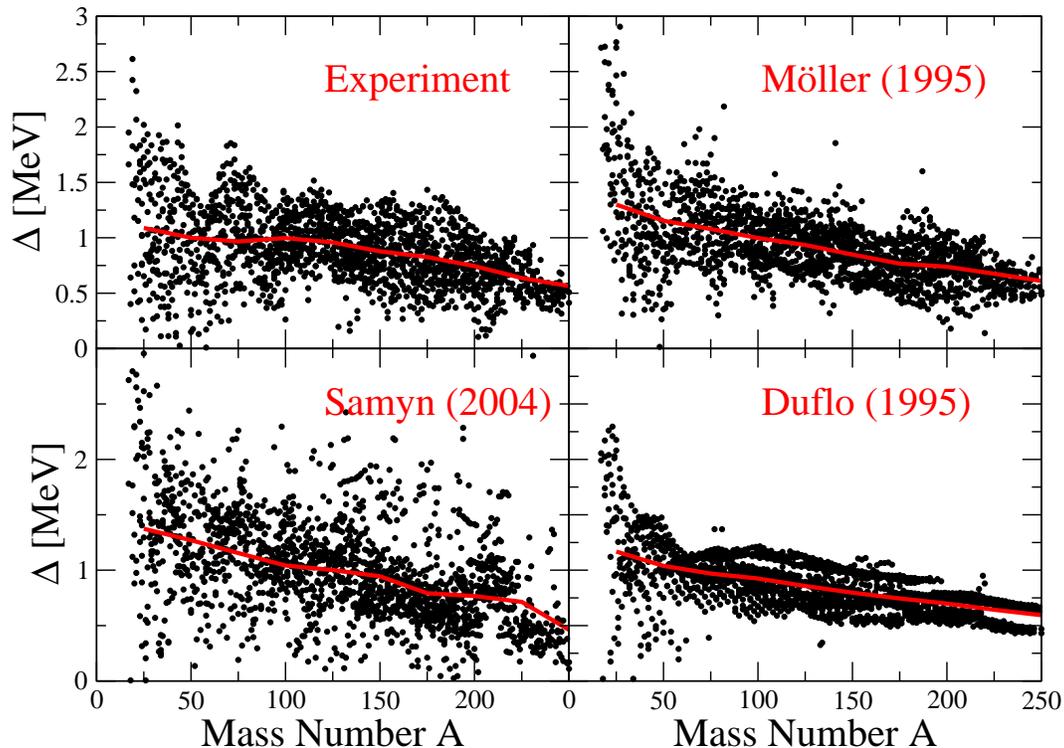}
\vspace{-0.3cm}
 \caption{Pairing gaps extracted from Eq.~(\ref{D3}) with odd
 $M$ versus mass number $A$ from experimental data, and three
 different theoretical mass models. Each black point corresponds
 to a nucleus. The solid red lines show the average pairing gap as shown in
 Fig.~\ref{Fig2}.}
 \label{Fig3}
\end{figure}
\begin{figure}
\includegraphics[width=10cm,clip=true]{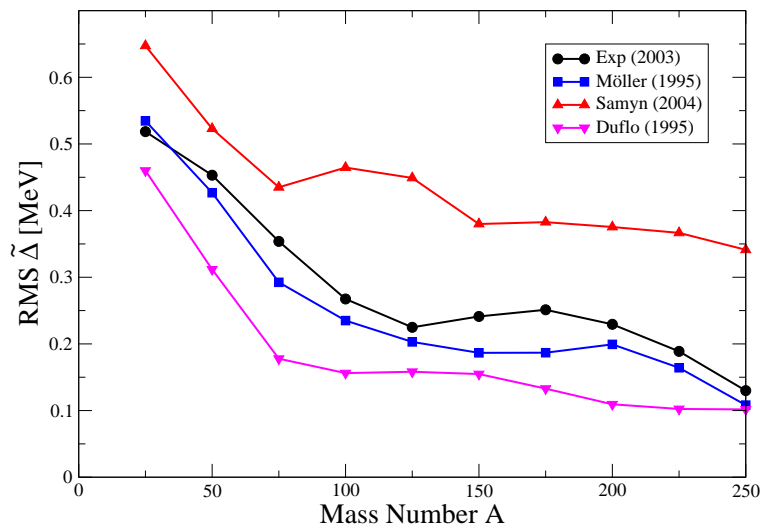}
\vspace{-0.3cm}
 \caption{Root mean square (RMS) values of pairing gaps $\Delta$ obtained from
 Eq.~(\ref{D3}) using odd $M$-values for three theoretical mass models,
 and experimental data. See caption of Fig.~\ref{Fig2} for notations.}
 \label{Fig4}
\end{figure}
The pairing gap as defined by Eq.~(\ref{D3}) can also be extracted
from theoretical mass models, and in Fig.~\ref{Fig2} we compare
the average value of $\Delta_3$ versus particle number for three
different mass models. Two of them are based on mean field theory.
The first one is a non-self-consistent macroscopic-microscopic
model \cite{Moller}, the second one is a self-consistent
calculation based on Skyrme-Hartree-Fock-Bogoliubov \cite{Samyn}
while the third one \cite{Duflo} is a shell--model based
calculation with parameterized monopole and multipole terms. Also
the experimental mean value of the pairing gap is shown in
Fig.~\ref{Fig2}, and it is seen that all three mass models give
similar results in good agreement with experimental numbers. If,
however, not average values, but pairing gaps obtained from all
nuclei are shown (Fig.~\ref{Fig3}), it is clear that the average
values give a poor description of the result. The fluctuations
(RMS value) of the pairing gaps are indeed very large, and exhibit
different variation with particle number for the different mass
models, see Fig.~\ref{Fig4}. All mass formula show the same
tendency of decreasing fluctuations with increasing mass number,
as is also seen from experimental masses. However, considerably
larger pairing fluctuations are seen in the mass model by Samyn et
al \cite{Samyn}, where, particularly for large mass numbers,
almost three times larger fluctuations are obtained, as compared
to pairing gap fluctuations obtained from measured masses. The
mass model based on the shell--model \cite{Duflo} gives
systematically too small fluctuations, while the mass model by
M\"{o}ller et al \cite{Moller} gives pairing fluctuations closest
to experimental data.

It is thus clear that the fluctuations of the pairing gap is a
most important property, and in the next section we shall present
a semiclassical theory for pairing fluctuations based on periodic
orbits \cite{Olofsson}.

\section{Periodic orbit description of pairing fluctuations}
The many-body Hamiltonian,
\begin{equation}
H=\hat{H}_1+\hat{V}_{\rm pair}=\sum_k e_k a^{\dagger}_k a_k -G
\sum_{k\ell} a^{\dagger}_k a^{\dagger}_{\bar{k}} a_{\bar{\ell}}
a_{\ell},
 \label{Htot}
\end{equation}
incorporates a one-body part, $\hat{H}_1$ (typically obtained from
a deformed mean-field), and a two-body pairing interaction between
time-reversed states, $\hat{V}_{\rm pair}$. All two-body matrix
elements are assumed to take the same value, $G$ (seniority
interaction). In the mean-field approximation in pairing space a
pairing gap, or pairing "deformation",
\begin{equation}
\Delta=\left< G\sum_k a^{\dagger}_k a^{\dagger}_{\bar{k}} \right>,
\end{equation}
is determined by the BCS gap equation \cite{BCS}
\begin{equation}
\frac{2}{G}=\sum_{k}\frac{1}{\sqrt{(e_{k}-\lambda)^2-\Delta^2}},
\end{equation}
that can be written as
\begin{equation} \label{gap1}
  \frac{2}{G}=\int_{-L}^L
  \frac{\rho(\eps)d\eps}{\sqrt{\eps^2+\Delta^2}},
\end{equation}
where $\rho(\eps)$ is the single--particle level density, and we
have put the Fermi energy, $\lambda$, to zero. The energy cut off
is at $\pm L$.

Following semiclassical approaches, the pairing gap as well as the
single-particle density of states are divided in a smooth part and
a fluctuating part, $\Delta=\Db+\Dt$ and $\rho=\rb+\rt$,
respectively. In the weak coupling limit $\Db \ll L$, the
smooth part of the gap is given by the well known solution $\Db =
2L \exp (-1/\rb G)$ (see e.g. Ref.~\cite{BCS}). In semiclassical
theory the fluctuating part of the density $\rt$ can be calculated
from purely classical properties \cite{semicl},
\begin{equation}
 \rt (\eps) = 2 \sum_p \sum_{r=1}^{\infty} A_{p,r} \cos(rS_p / \hbar + \nu_{p,r}),
\end{equation}
where the sum is over all primitive periodic orbits $p$ (and their
repetitions $r$) of the classical underlying effective
single-particle Hamiltonian, $H_1$. Each orbit is characterized by
its action $S_p$, stability amplitude $A_{p,r}$, period
$\tau_p=\partial S_p/\partial \eps$ and Maslov index $\nu_{p,r}$
(all evaluated at energy $\eps$). Assuming $\Dt \ll \Db$ and $\Db
\ll L$ gives after some algebra \cite{Olofsson}
\begin{equation} \label{dosc}
  \Dt =2 \frac{\Db}{\rb}\sum_p\sum_{r=1}^\infty A_{p,r}K_0(r\tau_p/\tD)\cos \left(
  \frac{rS_p}{\hbar}+\nu_{p,r} \right),
\end{equation}
where all classical quantities involved are evaluated at the Fermi
energy. $K_0(x)$ is the modified Bessel function of second kind,
and
\begin{equation}
\tD = \frac{h}{2\pi\Db}
\end{equation}
is a time corresponding to the pairing gap, that we may call the
pairing time. Since $K_0 (x) \propto \exp(-x)/\sqrt{x}$ for $x \gg
1$, the Bessel function exponentially suppresses all contributions
for times $\tau \gg \tD$ (making the sum in Eq.~(\ref{dosc})
convergent).

Since the value of the actions depend on the shape of the
mean--field potential, Eq.~(\ref{dosc}) predicts generically
fluctuations of the pairing gap as one varies, for instance, the
particle number, or the shape of the system at fixed particle
number. The fluctuations result from the interference between the
different oscillatory terms that contribute to $\Dt$. When the
motion is regular (integrable), continuous families of periodic
orbits having the same action, amplitude, etc, exist. The coherent
contribution to the sum (\ref{dosc}) of these families of periodic
orbits produces large fluctuations. In contrast, in the absence of
regularity or symmetries, incoherent contributions of smaller
amplitude coming from isolated unstable orbits are expected for
chaotic dynamics.

The second moment of the fluctuations may be obtained from
Eq.~(\ref{dosc}) as \cite{Olofsson}
\begin{equation} \label{var}
  \secmom{\Dt} = 2\frac{\Db^2}{\tH^2} \int_0^\infty d\tau K_0^2(\tau/\tD)
  K(\tau),
\end{equation}
where $\tH = h/\delta$ is the Heisenberg time ($\delta = \rb^{-1}$
is the single--particle mean level spacing at Fermi energy), and
$K(\tau)$ is the spectral form factor, i.e. the Fourier transform
of the two-point density--density correlation function
\cite{Berry}.

The structure of the form factor $K(\tau)$ is characterized by two
different time scales. The first one, the smallest of the system,
is the period $\tm$ of the shortest periodic orbit. The form
factor is zero for $\tau \leq \tm$, and displays non-universal
(system dependent) features at times $\tm \leqa \tau \ll \tH$. As
$\tau$ further increases, the function becomes universal,
depending only on the regular or chaotic nature of the dynamics,
and finally tends to $\tH$ when $\tau \gg \tH$. The result of the
integral (\ref{var}) thus depends on the nature of the dynamics,
and on the relative value of $\tD$ with respect to $\tm$ and
$\tH$. In the simplest approximation, all the non--universal
system--specific features are taken into account only through
$\tm$ \cite{Monastra}, and one can write $K(\tau) = 0$ for $\tau <
\tm$ and, for $\tau \geq \tm$, $K(\tau) = \tH$ for integrable
systems and $K(\tau) = 2\tau$ for chaotic systems with time
reversal symmetry. We assume a generic regular system; the analysis does not
apply to the harmonic oscillator, whose form factor is pathological.

This finally gives the expressions for fluctuations of the pairing
gap (normalized to the single--particle mean level spacing),
$\sigma=\rms{\Dt}/\delta$, assuming regular dynamics
\begin{equation}
  \sigma_{\rm reg}^2 = \frac{\pi}{4} \frac{\Db}{\delta} F_0 \left( D \right) \ ,
  \label{secmomreg}
\end{equation}
and assuming chaotic dynamics,
\begin{equation}
  \sigma_{\rm ch}^2 = \frac{1}{2\pi^2} F_1 \left( D \right) \ ,
    \label{secmomch}
\end{equation}
where we have introduced the function \cite{Olofsson}
\begin{equation}
F_n (D) = 1-\frac{\int_0^{D} x^n K_0^2(x) dx}{\int_0^{\infty} x^n
K_0^2(x) dx}.
 \label{F01}
\end{equation}
\begin{figure}
\includegraphics[width=10cm,clip=true]{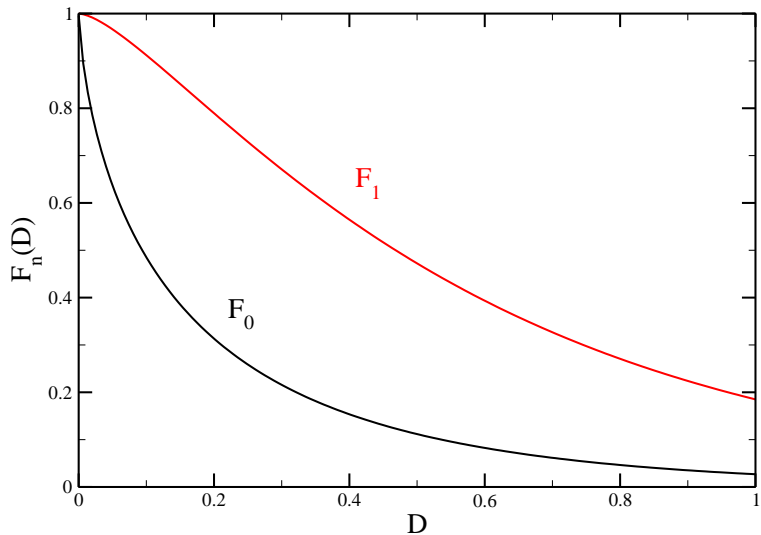}
\vspace{-0.3cm} \caption{The two functions $F_0$ and $F_1$, see
Eq.~(\ref{F01}), versus the argument $D$. } \label{Fig5}
\end{figure}
The argument $D$ is defined as
\begin{equation}
  D = \frac{\tm}{\tD}=\frac{2\pi}{g}\frac{\bar{\Delta}}{\delta},
  \label{D}
\end{equation}
where the parameter $g$ is the ratio between the Heisenberg time
and the time of the shortest periodic orbit,
\begin{equation}
g=\frac{\tau_H}{\tm}.
\end{equation}
This parameter is often called "dimensionless conductance". It
expresses the energy range, $L_{\rm max}(=g)$ (in dimensionless
units; energies are divided by the mean energy spacing, $\delta$),
in the single-particle spectrum where spectrum fluctuations show
universal properties \cite{Berry}. For a system where the
corresponding classical dynamics is regular/chaotic (with time
reversal symmetry), the statistical properties of the one-body
energies, $e_k$ of Eq.~(\ref{Htot}) are described by Poisson/GOE
(Gaussian Orthogonal Ensemble) statistics, respectively. The limit
$g\rightarrow \infty$ (i.e. $D$=0), corresponds to a universal
situation when the full spectrum corresponds to pure GOE (if
chaotic) or Poisson (if regular) statistics.

The dimensionless parameter $D$ can also be expressed as the
system size, $2R$, divided by the coherence length of the Cooper
pair, $\xi_0 = \hbar v_{\scriptscriptstyle F}/(2\Db)$, where
$v_{\scriptscriptstyle F}$ is Fermi velocity,
\begin{equation}
D=2R/\xi_0.
 \label{Dcoh}
\end{equation}
The Cooper pairs can thus be considered as restricted by the
system size if $D<1$.

In Fig.~\ref{Fig5} we show the two functions $F_0$ and $F_1$ of
Eq.~(\ref{F01}) versus the argument $D$. For small values of $D$,
$F_0\approx 1$ and $F_1\approx 1$, and a universal behavior
appears, i.e. the fluctuations do not depend on the system
properties, only the dynamics, implying Poisson and GOE statistics
for the full spectrum in the regular and chaotic cases,
respectively. In the other limit when $D$ becomes large, $F_0$ and
$F_1$ go to zero and all pairing fluctuations disappear. This is,
for example, the situation in bulk systems with a large number of
particles.
\begin{figure}
\includegraphics[width=16cm,clip=true]{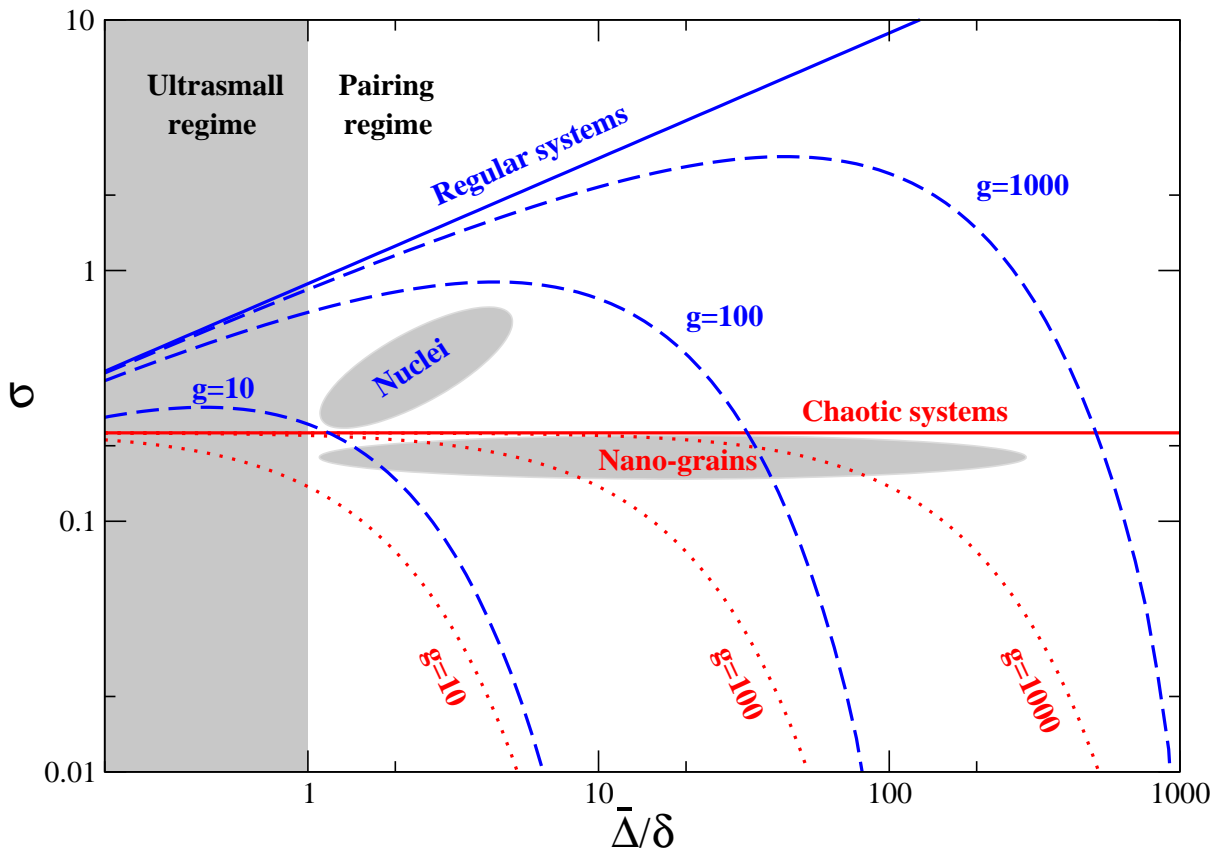}
\vspace{-0.3cm}\caption{Fluctuations of the pairing gap as a
function of the mean
  value for mesoscopic systems (log-log scale; all quantities normalized with the
  single-particle mean spacing). Regular and chaotic dynamics are
  shown by blue and red lines, respectively. The dashed curves
  correspond to different values of the dimensionless conductance,
  $g$, and the limiting case of $g \rightarrow \infty$ is shown by
  solid lines. The results are valid in the pairing regime
  $\bar{\Delta}/\delta > 1$. Applications to Nuclei and
  Nano-grains are marked out. Ultracold atomic gases may be controlled to
  appear in major parts of the figure. From Ref.~\cite{Olofsson}.}
  \label{Figgeneral}
\end{figure}

In Fig.~\ref{Figgeneral} pairing gap fluctuations are shown versus
the pairing gap for different values of the dimensionless
conductance, $g$, for regular as well as for chaotic dynamics. The
plot covers a large range of parameter values and is shown in
log-log scale. Pairing gaps of the order of the mean level spacing
or smaller, $\bar{\Delta}<\delta$, (ultrasmall regime), are not
treated by the present theory and corresponding region is shaded
in the figure. In this region the Anderson condition
\cite{Anderson} implies no BCS pairing. As mentioned above,
non-universal behavior (deviating from GOE or Poisson) appears
when $D>0$, corresponding to finite values of the dimensionless
conductance, $g$.

\section{Fluctuations of pairing gap in nuclei}

In a previous section we studied pairing gap fluctuations in
nuclei as obtained from the odd-even mass difference, where
binding energies were taken from different mass models as well as
from data, see Fig.~\ref{Fig4}. We now would like to use the
semiclassical theory developed in previous section to compare
pairing gap fluctuations, and compare to these results. Notice
that the theory does not contain any parameters. Once the system
is defined, and the variable $D$ has been determined, the
fluctuations only depend on the corresponding classical dynamics,
being regular or chaotic. In nuclei ground states the dominating
dynamics is expected to be regular, although elements from chaos
may be present, see Ref.\cite{masses}.

\begin{figure}
\includegraphics[width=10cm,clip=true]{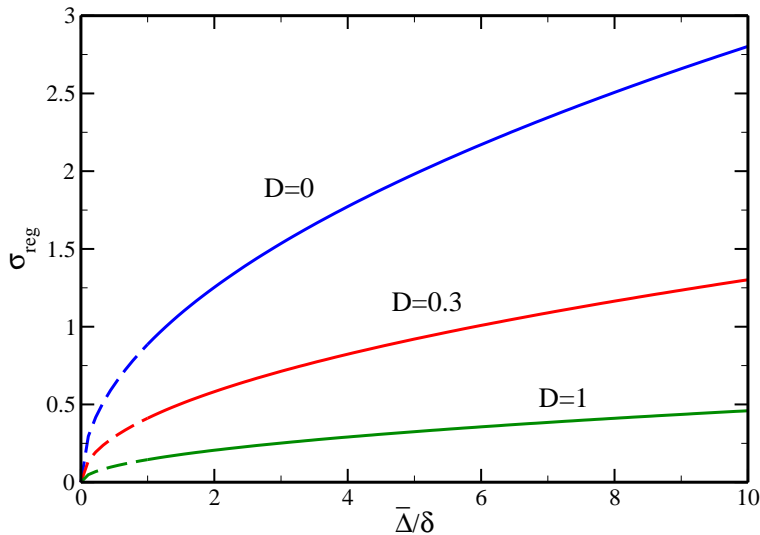}
\vspace{-0.3cm}
 \caption{Fluctuations of pairing gap versus average pairing gap
 for different values of $D$. Regular dynamics is assumed.
 $D=0.3$ approximately corresponds to nuclei. Fluctuations
 and pairing gaps are expressed in units of the mean level spacing, $\delta$. }
 \label{Fig7}
\end{figure}
To evaluate the root mean square value (RMS) of the pairing
fluctuations from the theoretical expressions,
Eqs.~(\ref{secmomreg}) and (\ref{secmomch}), we only have to
determine the parameter $D$. The size of the nucleus,
$2R=2\times 1.2A^{-1/3}$ fm, and size of the pairing correlation length,
$\xi_0=\hbar v_F/(2\Db)=11.3A^{-1/4}$ fm, give
(using Eqs.~(\ref{Dbeq}) and (\ref{Dcoh}))
$D=2R/\xi_0=0.22A^{1/12}=0.27-0.33$ for mass numbers in the
interval, $A$=25-250. The dependence of the pairing fluctuations,
assuming regular dynamics, with the (average) pairing gap is shown
in Fig.~\ref{Fig7} for the three cases, $D$=0, 0.3 and 1. If the
pairing correlation length is smaller than the system size ($D>1$)
pairing fluctuations are quite small. Largest fluctuations appear
in the universal limit when $D$=0. The small values of $D$ for
nuclei ($D \approx$ 0.3) implies that the Cooper pairs are
non-localized. The pairing gap fluctuations are thus substantial,
and are about half the value at the universal limit.

\begin{figure}
\includegraphics[width=10cm,clip=true]{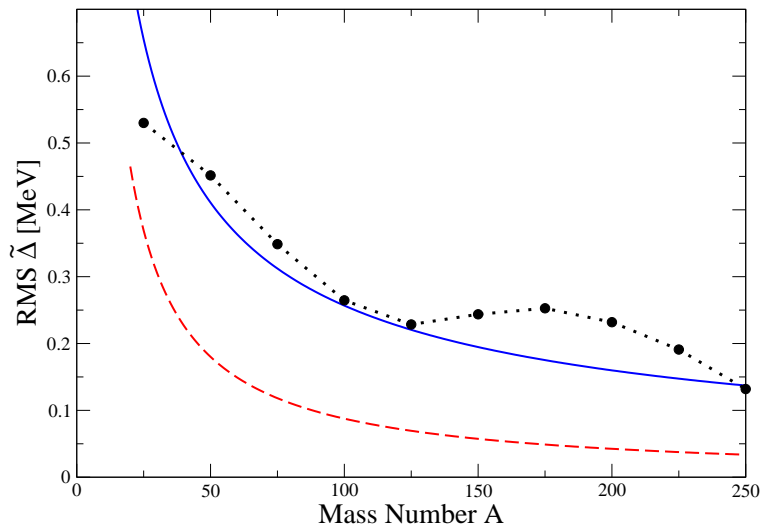}
\vspace{-0.3cm}
 \caption{RMS values of pairing gap fluctuations in
nuclei versus mass number $A$, obtained from measured masses
(dashed line connected by filled circles). Semiclassical
calculations of the fluctuations assuming chaotic
(Eq.~(\ref{secmomch})) and regular (Eq.~(\ref{secmomreg}))
dynamics are shown by red dashed and blue solid lines,
respectively. From Ref.~\cite{Olofsson}}
 \label{Fig9}
\end{figure}
We may compare the fluctuations to the experimental pairing gap
fluctuations by inserting the above values for $D$, and the
average values of $\Delta$ from Eq.~(\ref{Dbeq}), in
Eqs.~(\ref{secmomreg}) and (\ref{secmomch}), assuming regular or
chaotic dynamics. The resulting curves are compared to the
experimental one in Fig.~\ref{Fig9}. Note the good agreement
between the theoretical pairing gap fluctuations assuming regular
dynamics, and the experimental curve, both in the overall
amplitude and in the $A$--dependence. In Ref.\cite{masses} it was
discussed the possibility that the dynamics of the nuclear ground
state is mixed regular and chaotic. Making this assumption in the
calculation of fluctuations of the pairing gap results in a curve
that is very close to the purely regular curve in Fig.~\ref{Fig9}.
That is, fluctuations of the pairing gap cannot distinguish if
there is a chaotic component in the nuclear ground state.

\section{Shell structure in pairing gap from periodic orbit
theory}
\begin{figure}
\includegraphics[width=7.3cm,clip=true]{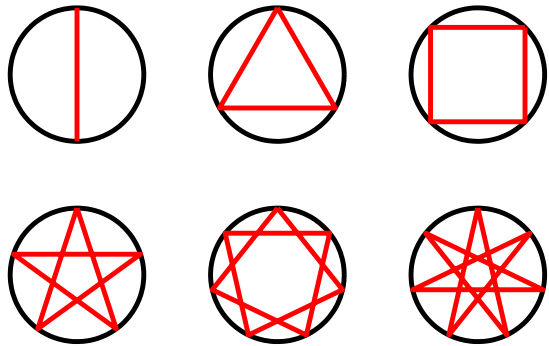}
\vspace{-0.3cm}
 \caption{A sample of classical periodic orbits of the cavity potential. The
   orbits shown correspond to the index $(v,w)$=(2,1), (3,1), (4,1), (5,2),
 (7,2) and (7,3), counted from upper left to lower right figure.}
  \label{Fig10}
\end{figure}
\begin{figure}
\includegraphics[width=14cm,clip=true]{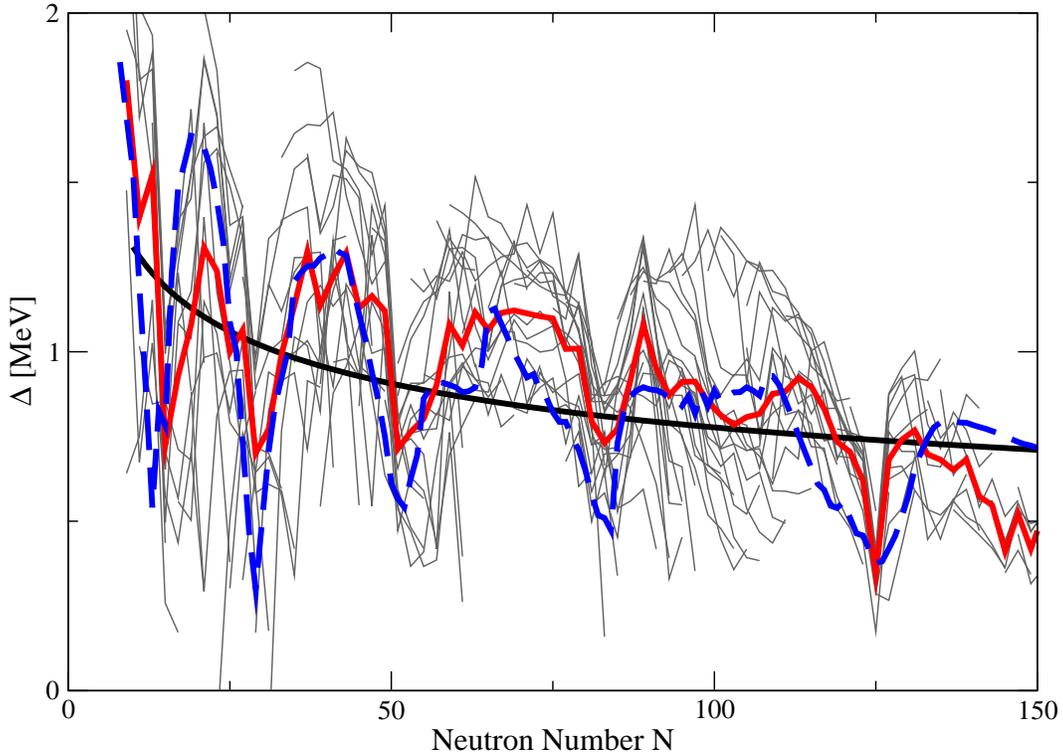}
\vspace{-0.3cm} \caption{Nuclear pairing gaps for neutrons.
Experimental isotope sequences are connected by thin solid lines.
The blue dashed line shows average pairing gaps. Calculations from
the cavity model are shown by the solid red line. Average behavior
(Eq.~(\ref{Dbeq})) is shown by solid black line. Data from
Ref.~\cite{NuclMass}.}
  \label{Fignucl}
\end{figure}
Shell effects in nuclei are also seen in the pairing gap. One may
go beyond a statistical description, and use Eq.~(\ref{dosc}) to
obtain a detailed description of the variation with neutron or
proton numbers. For that purpose, we assume for the nuclear mean
field a simple hard-wall cavity potential. The shape of the cavity
at a given number of nucleons is fixed by minimization of the
energy against quadrupole, octupole and hexadecapole deformations
\cite{Hasse}. To simplify, we take $N=Z$. The periodic orbits of
the spherical cavity (a few short orbits are shown in
Fig.~\ref{Fig10}) are used in Eq.~(\ref{dosc}), with modulations
factors that take into account deformations and inelastic
scattering \cite{creagh}. This gives
\begin{equation}
\tilde{\Delta}=\frac{\bar{\Delta}}{\bar{\rho}E_0} \sum_{v,w} A_{v
w}M_{v w}(x)\kappa_\chi(\ell_{v w})K_0 \left(\frac{\ell_{v
w}\bar{\Delta}}{2\bar{k_F}RE_0} \right) sin(\bar{k_F}R \ell_{v w}
+ \nu_{v w} \frac{\pi}{2}),
\end{equation}
where $M_{v w}(x)$ is a modulation factor for perturbative
deformations, $\kappa_\chi (\ell_{v \omega})$ a modulation factor
for inelastic scattering, and $x$ stands for the three considered
deformation degrees of freedom, quadrupole ($\varepsilon_2$),
octupole ($\varepsilon_3$) and hexadecapole ($\varepsilon_4$). The
summation is carried out over the two indices $(v,w)$ including the periodic
orbits shown in
Fig.~\ref{Fig10}. We set the average of $\Dt$ to zero, as was done
with the experimental data. In Fig.~\ref{Fignucl} we compare the
theoretical result $\Dt (N)$ to the experimental value averaged
over the different isotopes at a given $N$. The agreement is
excellent; the theory describes all the main features observed in
the experimental curve.

\section{Pairing fluctuations in nano-sized metallic grains}
Experiments in the 90's have explored the superconducting
properties of nanometer scale aluminum grains \cite{RBT}.
Irregular shape of the grains implies chaotic dynamics, and energy
levels are found to follow GOE (there are no further symmetries
than time-reversal), see Ref.~\cite{NanoRev}. The existence of a
superconducting gap was demonstrated in the regime $\Db > \delta$,
whereas no gap was observed when $\Db < \delta$. The transition
occurs around $N \sim 5000$, where $N$ is the number of conduction
electrons in the grain. The $N$ dependence of the average gap
$\Db$ is poorly understood. We will adopt for grains the thin-film
value $\Db \approx 0.38 \times 10^{-3}$~eV \cite{NanoRev}. The
mean level spacing is $\delta = (2\ef)/(3N) \approx 2.1/N$~eV,
whereas the dimensionless conductance, $g \approx 2.6 N^{2/3}$.
Eq.~(\ref{D}) gives $D \approx 4.4 \times 10^{-4} N^{1/3}$, which
ranges from $0.05$ to $0.02$ when $N$ varies between $10^{3}$ and
$10^{5}$. This means that the variance will be close to its
``universal'' value (GOE limit) obtained by setting $F_1 (D) = 1$
in Eq.~(\ref{secmomch}), namely \cite{Comm}
\begin{equation}
\sigma^2_{\rm ch} = 1/2 \pi^2.
\end{equation}
The typical range of variation of pairing gap fluctuations for
nano-grains is marked out by a grey area in Fig.~\ref{Figgeneral}.

There are no explicitly measured pairing gap fluctuations for
nanosized metallic grains to compare the present theory. We may,
however, compare to other independent calculations, namely the
condensation energy calculated of chaotic grains \cite{Dukelsky},
and variation of the pairing gap with particle number in a regular
cubic shaped grain \cite{Garcia}.

The BCS condensation energy is defined as the total energy
difference between the paired and the unpaired system. With our
choice of gap equation (\ref{gap1}) it is written as
\begin{equation}
  E_{\rm C} = E_{\rm tot}(\Delta)-E_{\rm tot}(\Delta=0) =
  \int_{-L}^L \rho(e) e v^2(e) de -\frac{\Delta^2}{4G}-\int_{-L}^0 \rho(e) e dx
\end{equation}
where $v^2(e)=\frac{1}{2}\left(1-\frac{e}{\sqrt{e^2+\Delta^2}}
\right)$. Inserting the semiclassical approximations of
$\rho=\rb+\rt$ and $\Delta=\Db+\Dt$ and expanding $E_{\rm
C}=\Ecb+\Ect$ to lowest order in fluctuating properties,
assuming $\Ect \ll \Ecb$, gives
\begin{equation}
\Ect = -2\sum_p \sum_{r=1}^\infty A_{p,r} \left[Q_{p,r}
  +qK_0(r\tau_p/\tD) \right]
  \cos\left(\frac{rS_p}{\hbar}+\nu_{p,r}\right)
\end{equation}
where $K_0$ is the modified Bessel function of second kind,
$q=\Db^2\left(1-\frac{1}{2\rb G} \right)$ and
\begin{equation}
Q_{p,r}=\int_{-L}^0 \cos\left(\frac{r\tau_p}{\hbar}\right)e
\left(1+\frac{e}{\sqrt{e^2+\Db^2}} \right) de
\end{equation}

The second moment, $\sigma_{\rm C}^2=\left< \Ect^2 \right>$, is
thus obtained as
%\begin{align}
%  \left< \Ect^2 \right> = &\frac{2}{h^2} \int_{\tm}^\infty
%(\tau)Q^2(\tau)d\tau+\nonumber\\
%  +  &\frac{4q}{h^2} \int_{\tm}^\infty K(\tau)Q(\tau) K_0(\tau) d\tau+ \nonumber\\
%  + &\frac{2q^2}{h^2} \int_{\tm}^\infty K(\tau)K_0^2(\tau) d\tau
%\end{align}
\begin{equation}
  \sigma_{\rm C}^2 = \frac{2}{h^2} \int_0^\infty \left[Q(\tau)+qK_0(\tau)
  \right]^2 K(\tau) d\tau,
  \label{sigmac}
\end{equation}
where $K(\tau)$ is the spectral form factor.
%\begin{equation}
%\sigma_{\rm C} = \sqrt{\left<\Ect^2 \right>}
%\label{sigmac}
%\end{equation}
\begin{figure}
\includegraphics[width=10cm,clip=true]{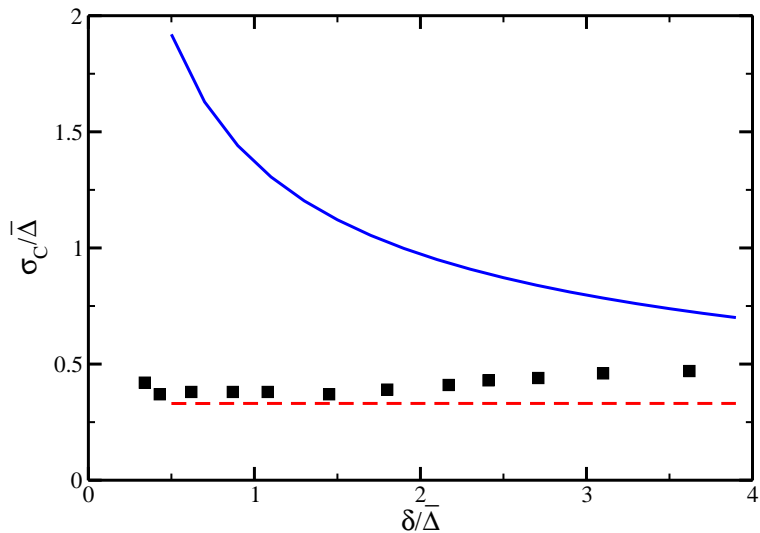}
\vspace{-0.3cm}
 \caption{Fluctuations of the condensation energy as a function of the mean
   level spacing of the system. The blue solid and red dashed lines show
   the numerical calculations of
   regular and chaotic fluctuations according to Eq.~(\ref{sigmac}). The
   squares show the calculations of ref.~\cite{Dukelsky}.}
 \label{ectilde}
\end{figure}
Using in addition the estimates $G=\frac{0.224}{\rb}$ and
$L=\frac{\Db}{2}e^{1/0.224}$ \cite{Dukelsky} we show in
Fig.~\ref{ectilde} the fluctuations of the condensation energy,
$\sigma_{\rm C}$, versus the level distance in the system,
assuming regular and chaotic motion. As mentioned, there is no
experimental data to compare to, but we may compare to a numerical
calculation of the $\sigma_{\rm C}$ by Sierra {\it et.
al.}~\cite{Dukelsky} where they use the Richardson's solution of
the pairing problem and random matrix theory (GOE) for generating
the spectrum. We see that our semiclassical theory, assuming
chaotic dynamics, agrees well with the random matrix calculation
of Ref.~\cite{Dukelsky}. The random matrix model applied in
\cite{Dukelsky} happens to be a reasonable approximation for the
considered nano-grains, since $D\approx 0$, i.e. the nano-grains
have properties which are close to the universal limit.

Garcia-Garcia et al \cite{Garcia} studied shell effects on the
pairing gap in a cubic geometry as a function of the mean pairing
gap, $\Db/\delta$.
%The studies were performed around
%$\Delta/\delta$=5 and $\Delta/\delta$=12.
The cubic geometry implies classically regular motion, and we may
thus use Eq.~(\ref{secmomreg}) above to calculate the fluctuation
of the paring gap, normalized to the mean pairing gap $\Db$ instead of the
mean level spacing $\delta$,
$\sigma_{\rm reg}\delta/\bar{\Delta}=\sqrt{\frac{\pi}{4}\frac{\delta}{\bar{\Delta}}}$,
%giving 0.40 and 0.26, respectively,
where we have set $F_0=1$ since for small grains $D\approx 0$. This gives a
very good agreement to the pairing fluctuations calculated in
Ref.~\cite{Garcia}. It is interesting to note that by making the
system chaotic, the pairing fluctuations decreases substantially
and become (see Eq.~(\ref{secmomch})),
$\sigma_{\rm ch}\delta/\bar{\Delta}=\frac{1}{\sqrt{2}\pi}\frac{\delta}{\bar{\Delta}}$,
i.e.
$\sigma_{\rm ch}/\sigma_{\rm reg}=\sqrt{\frac{2}{\pi^3}\frac{\delta}{\bar{\Delta}}}\approx0.05$
if $\bar{\Delta}/\delta=20$.

\section{Pairing fluctuations in ultracold fermionic gases}
Recently, a large interest has emerged in studying trapped atomic
gases of bosons and fermions. The gases are ultracold and dilute,
and provide the possibility to study new quantum phenomena in the
physics of finite many-body systems. The number of neutrons of the
confined atoms determines the quantum statistics; odd number
implies Fermions, and even number implies bosons. Studies of Bose
condensates can be done for the bosonic gases, and studies of
quantum phenomena including superfluidity and the transition to a
Bose-Einstein condensate can be conducted for the Fermi gases.

Since the atom-atom interaction is short ranged and much smaller
than the interparticle distance, the atom-atom interaction can be
approximated by the $\delta$-interaction,
\begin{equation}
V(r_1-r_2)=4\pi\frac{\hbar^2a}{m}\delta^{(3)}(r_1-r_2),
\end{equation}
where $a$ is the s-wave scattering length, that can be externally
controlled in size and even in sign through the Feshbach
resonance. Also the confinement potential can be externally
controlled to create regular as well as chaotic dynamics.
Ultracold fermionic gases thus provide excellent conditions for
theoretical as well as experimental studies of pairing properties,
including pairing fluctuations. Since both particle number and
interaction strength are experimentally controlled parameters, the
fluctuations may appear in major parts of Fig.~\ref{Figgeneral}.

We estimate $\delta = (2\ef)/(3N)$ and $g=\frac{1}{2}(3N)^{2/3}$;
in the dilute BCS region $\Db=(2/e)^{7/3} \ef \exp\left( -\pi/2
\kf |a| \right)$ \cite{Gorkov}, with $\kf$ the Fermi wavevector,
giving
\begin{equation} D=\frac{2R}{\xi_0}=2\pi (2/e)^{7/3}
(3N)^{1/3} \exp \left(-\pi/2\kf |a| \right).
\end{equation}
Recent experiments using Li$^6$ reach $\kf |a| = 0.8$
\cite{KetterleLi6}, implying negligible fluctuations for typical
values of $N \sim 10^6$. Reducing to $\kf |a|=0.2$ and $N =
10^4$ yields for generic regular systems fluctuations that are on
the same magnitude as the mean pairing gap, $\sigma_{\rm reg}
\approx 0.5\Db / \delta$.

\section{Summary}
In summary, we have presented a semiclassical theory that provides
a generic description of fluctuations and shell structure of the
BCS pairing gap in finite Fermi systems. These mesoscopic systems
are generically dominated by system specific features not included
in purely statistical models like GOE. Different possible regimes,
as well as the influence of order/chaos dynamics, were
investigated, in particular for the typical size of the
fluctuations (Fig.~\ref{Figgeneral}). The present theory provides
analytic predictions, valid for a wide range of physical
situations. It also compares very favorably with available
experimental data.

\vspace{1cm}
S.\AA. thanks the Swedish Science Research Council,
and P.L. acknowledges support by grants ANR--05--Nano--008--02,
ANR--NT05--2--42103 and by the IFRAF Institute.


\begin{thebibliography}{99}
\bibitem{Comm} The expression for pairing fluctuations in the universal GOE limit
(corresponding to $D$=0) was first derived in \cite{ML}.
\end{thebibliography}
\end{document}